\documentclass[prb,twocolumn,floatfix,superscriptaddress,showpacs]{revtex4}
\usepackage{graphicx}
\usepackage{amsmath,amssymb,amsfonts,bm}
\usepackage{hyperref}

\begin{document}
\title{Spin-polarized currents through interacting quantum wires with nonmagnetic leads}
\author{J.E.~Birkholz} 
\affiliation{Institut f\"ur 
Theoretische Physik, Universit\"at G\"ottingen, 
Friedrich-Hund-Platz 1, D-37077 G\"ottingen, Germany}
\author{V.~Meden}
\affiliation{Institut f\"ur Theoretische Physik A, RWTH Aachen University and 
JARA---Fundamentals of Future Information Technology, D-52056 Aachen, Germany}

\begin{abstract}
We study the performance of a quantum wire spin filter that is based on the Rashba spin-orbit 
interaction in the presence of the electron-electron interaction. The finite length wire 
is attached to two semi-infinite nonmagnetic leads. Analyzing the spin polarization of the 
linear conductance at zero temperature, we show that spin-filtering is possible by 
adequate tuning of the system parameters first considering  noninteracting electrons. 
Next, the functional renormalization group method is used to capture correlation effects 
induced by the Coulomb interaction. For short wires we show that the energy regime in which 
spin polarization is found is strongly affected by the Coulomb interaction. For long wires 
we find the power-law suppression of the total conductance on low energy scales typical 
for inhomogeneous Luttinger liquids while the degree of spin polarization stays constant.  
\end{abstract}
\maketitle

\section{Introduction}
\label{intro}
The birth of \textit{spintronics} can be dated back to the discovery of the \textit{giant 
magnetoresistance} effect in 1988.\cite{baibich, binasch} Since then, many theoretical and 
experimental studies on spin-dependent electronic transport have been performed in order to 
achieve a clear understanding of the underlying physics and to investigate the possibility 
of fabricating spintronic devices.\cite{datta,egues,beenakker} One manifest realization 
of such a device is a quantum wire with parameters which can be tuned by a set of gate 
electrodes such that the transport of electrons with a certain spin direction is favored.
The original proposal by Datta and Das\cite{datta} is to use the spin 
precession in a narrow-gap semiconductor wire with spin-orbit coupling between two magnetized 
leads to modulate the current. However, Str\v{e}da and S\v{e}ba came up with 
an idea of obtaining spin polarization in a (almost) nonmagnetic system. They considered transport 
through an {\it infinite} quasi one-dimensional (1D) quantum wire with Rashba spin-orbit interaction (SOI), 
a magnetic field in the direction of the propagation as well as a potential 
step.\cite{streda} In their setup, leads with negligible SOI and 
vanishing magnetic field as present in any experimental realization were not taken into 
account. 

The goal of the present work is twofold. We {\it first} investigate, under which conditions spin-polarized 
currents through a {\it finite length} 1D wire with SOI and parallel magnetic field can be achieved, 
if the coupling to two semi-infinite {\it leads} is included. We study the influence of a potential 
step and a localized impurity. Such inhomogeneities play an important role for the spin polarization. 

In the Datta-Das setup as well as in the system suggested by Str\v{e}da and S\v{e}ba the 
spin dependent transport properties heavily rely on the strictly 1D nature of the quantum wire with 
only one partially filled subband. Already in the presence of a few filled subbands the control 
over the spin ceases.\cite{Uli} In a few hundred nanometers long strictly 1D wires the 
Coulomb repulsion of 
electrons is expected to have a dramatic effect on the physics. Such systems cannot be 
described by Fermi liquid 
theory. Instead, the low-energy properties are 
captured by the {\it Luttinger liquid} phenomenology.\cite{KS} 
One can thus expect that electron correlations will also affect the performance of 
spintronic devices made of 1D wires.\cite{haeusler,Uli,iucci,mikhail} In particular, local 
inhomogeneities, which are important to achieve controllable spin polarization in 
the Str\v{e}da and S\v{e}ba setup, strongly suppress the conductance in Luttinger 
liquids.\cite{KaneFisher,Furusaki} It is thus mandatory to investigate how the spin 
polarization is affected by Luttinger liquid physics. This is the {\it second} goal of our 
work. A first step in analyzing the performance of the  Str\v{e}da and S\v{e}ba spin 
filter in the presence of the electron-electron interaction was taken in Ref.~\onlinecite{Martin}. 

Standard methods such as the self-consistent Hartree-Fock approximation do not capture the 
Luttinger liquid physics of our setup and are known to 
lead to severe artifacts if being applied to low-dimensional electron systems with Coulomb 
interaction. We therefore use an approximation which is based on the functional renormalization 
group (fRG) approach to treat the two-particle interaction in a model Hamiltonian.  In the absence 
of SOI, it was shown to be a reliable tool to calculate the linear conductance of inhomogeneous 
quantum wires for weak to intermediate interactions.\cite{enss,andergassen1} 

We show that for vanishing two-particle interaction spin polarized currents can be 
obtained using a similar mechanism as in 
Ref.~\onlinecite{streda}, even in the presence of nonmagnetic leads with vanishing SOI 
and zero magnetic field. 
We then include the Coulomb interaction in our analysis. We first consider short wires (several tens 
of nanometers) in which Luttinger liquid physics does not become apparent and investigate how the 
spin polarization is affected by the two-particle interaction. We find that it strongly modifies 
the energy regime (energy of the incoming particles) in which spin polarization can be 
achieved. Next, the focus is on system sizes for which Luttinger 
liquid behavior is apparent in the absence of SOI  (wires of several hundreds of nanometers). 
In particular, we analyze if the spin polarization of 
an inhomogeneous wire is suppressed as a function of an 
infrared energy scale. We find that although the total conductance shows a power-law suppression 
in the presence of SOI (similar to the situation with vanishing SOI) the polarization {\it does not} 
follow such a scaling law. We present indications that the degree of spin polarization in the 
presence of Coulomb interaction might even exceed the one obtained for vanishing 
two-particle interaction.   

While in most studies on spintronic devices the correlations are 
neglected even if the suggested setups contain 1D quantum wires our results clearly 
reveal the importance of the two-particle Coulomb interaction in the spin filter 
suggested in Ref.~\onlinecite{streda}.  We here refrain from making direct 
contact to 
existing or future experiments as we are mainly interested in studying the basic physics within a 
simplified model. It must certainly be extended to be considered realistic. However, our 
parameters are taken from a physically sensible range (see below).       

This paper is organized as follows. In the next section, we introduce our setup and lattice model. The 
techniques to obtain the linear conductance are described in Sect.~\ref{method}. In Sect.~\ref{results}, 
we first present our results for spin-polarized transport at vanishing two-particle interaction 
in the presence of leads. We next include the Coulomb interaction and study 
its interplay with the SOI, the magnetic field, and external potentials. Our results are summed up 
in Sect.~\ref{summary}.

\section{The model}
\label{theory}
\subsection{Spin-orbit interaction}

The two prototypical experimental systems for strictly 1D quantum transport are carbon nanotubes
and confined electron gases which form at the interface of properly designed semiconductor heterostructures.
In the latter, the confining potential generically leads to a sizable SOI and these are the type of 
systems we have in mind in the following. But also in the former the SOI seems to be surprisingly 
large.\cite{Kuemmeth} We choose our 
coordinate system such that the heterostructure confines the two-dimensional electron gas 
in $z$-direction and the external potential results in a confinement of the electrons 
in $y$-direction. Therefore, the electrons are able to move in $x$-direction only. Since 
we assume the confinement to be very sharp, the different electronic subbands will be well 
separated. For a sufficiently low electron density, we can thus focus on the lowest 
subband and neglect any subband mixing. 

The sharp confinement leads to large electric fields, 
which induce a spin-orbit coupling\cite{winkler} 
\begin{equation}
H_\mathrm{SO}=-\frac{e\hbar}{4m^2 c^2} \mbox{\boldmath{$\sigma$}} \cdot 
\left[\mathbf{E}\times\left(\mathbf{p}-\frac{e}{c}\mathbf{A}\right)\right]
\label{H_SO}
\end{equation} 
with the electric field $\mathbf{E}=-\nabla V/e$ ($e<0$ is the
electron charge) being the gradient of the ambient potential. For the 1D infinite 
noninteracting continuum model (1D electron gas), SOI results in a horizontal splitting of the quadratic 
electron energy dispersion $\epsilon(k,s)$ with wave number $k$ and $s=\pm$ 
being an additional quantum number. 
Within a certain parameter regime 
a magnetic field in $x$-direction, which couples to the electron spin, leads to a deformation of the
lower parabolic branch resulting in a double well form (see the low-energy region of the 
two central dispersions  of Fig.~\ref{system_leads}). 
Moreover, it was shown in 
Ref.~\onlinecite{streda} and extensively discussed in Ref.~\onlinecite{birkholz1} that 
the spin expectation values become $k$-dependent and display a rich behavior in the presence of
both SOI and magnetic field. Applying an additional step-like potential and considering 
electrons with energy in regions of only double degeneracy leads to the above mentioned
spin-polarized transport.\cite{streda,birkholz1}

\subsection{Lattice system}

The starting point of our investigation is a 1D noninteracting tight-binding lattice Hamiltonian. 
As our single-particle basis, we choose Wannier states $|j,\sigma\rangle$ with
$j\in\{1,...,N\}$ labeling the lattice site and
$\sigma=\uparrow,\downarrow$ denoting the spin. 
The spin quantization is chosen along the $z$-direction. The Hamiltonian can be written as
\begin{equation}
H_0=H_\mathrm{free}+H_\mathrm{pot}+H_R+H_Z \, ,\label{H0}
\end{equation}
with the free part
\begin{equation}
H_\mathrm{free}=-t\sum_{j=1}^{N-1} \sum_\sigma 
\left(c_{j+1,\sigma}^\dagger c_{j,\sigma} + c_{j,\sigma}^\dagger  
c_{j+1,\sigma}\right) \, ,
\end{equation}
describing the hopping of amplitude $t>0$ and the external potential (e.g.~due to nano-device structuring)
\begin{equation}
H_\mathrm{pot}=\sum_{j=1}^N \sum_\sigma V_{j,\sigma} 
c_{j,\sigma}^\dagger c_{j,\sigma} \, .\label{H_pot}
\end{equation}
Here, $c_{j,\sigma}^\dagger$ denotes the creation operator of an electron
at site $j$ with spin $\sigma$.
In a lattice model the effect of SOI due to confinement in $z$-direction is accounted 
for by a spin-flip hopping,\cite{mireles} and the effect of SOI due to confinement 
in $y$-direction by an imaginary spin-conserving hopping.\cite{birkholz1} Thus the SOI (Rashba) term reads 
\begin{eqnarray}
H_R&=&-  \sum_{j=1}^{N-1}  \sum_{\sigma,\sigma'} \alpha_{z,j} \left( c_{j+1,\sigma}^\dagger
\left(i\sigma_y\right)_{\sigma,\sigma'} c_{j,\sigma'} + \mbox{H.c.}
\right)
\\ \nonumber
&&+\sum_{j=1}^{N-1} \sum_{\sigma,\sigma'} \alpha_{y,j} \left( c_{j+1,\sigma}^\dagger
\left(i\sigma_z\right)_{\sigma,\sigma'} c_{j,\sigma'} + \mbox{H.c.} \right) \, .
\end{eqnarray}
The SOI coupling constants $\alpha_{z,j}>0$ and $\alpha_{y,j}>0$ are assumed to depend 
on the bond $(j,j+1)$ considered.\cite{remarkfootnote}
The effect of a magnetic field is captured by the Zeeman term 
\begin{eqnarray}
H_Z&=& \gamma \sum_{j=1}^{N} \sum_{\sigma,\sigma'} B_j  c_{j,\sigma}^\dagger\left[
  \left(\sigma_x\right) 
_{\sigma,\sigma'}\sin\theta\cos\varphi \right. \label{eqnzeeman} \\ \nonumber
&& + \left. \left(\sigma_y\right) 
_{\sigma,\sigma'}\sin\theta\sin\varphi+\left(\sigma_z\right) 
_{\sigma,\sigma'}\cos\theta \right]  c_{j,\sigma'} \, ,
\end{eqnarray}
with a site dependent magnetic 
field in (for now) arbitrary direction
$\mathbf{B}_j=B_j \left(\sin\theta\cos\varphi,\sin\theta\sin\varphi,\cos\theta\right)$,
and $\gamma$ being the Zeeman coupling constant.
For $N \to \infty $ this lattice
model was shown to give a similar  low-energy dispersion as the continuum model as 
well as a similar energy dependence of the 
spin expectation values.\cite{birkholz1} 

We supplement our model by a 
local site-dependent Coulomb interaction $U_{1,j}$ 
\begin{equation} 
H_1=\sum_{j=1}^{N} \sum_{\sigma,\sigma'} U_{1,j} c_{j,\sigma}^\dagger c_{j,\sigma}
  c_{j,\sigma'}^\dagger c_{j,\sigma'}(1-\delta_{\sigma,\sigma'})\label{H1}
\end{equation}
and a bond-dependent nearest-neighbor Coulomb interaction $U_{2,j}$ 
\begin{equation}
H_2=\sum_{j=1}^{N-1} \sum_{\sigma,\sigma'} U_{2,j} c_{j+1,\sigma}^\dagger c_{j+1,\sigma} 
  c_{j,\sigma'}^\dagger c_{j,\sigma'} \label{H2}
\end{equation}
(extended Hubbard model). The total Hamiltonian of the quantum wire is given by
\begin{equation}
H=H_0+H_1+H_2. \label{totalham}
\end{equation}

At the sites $1$ and $N$ the 1D wire is (end-)coupled to two semi-infinite 
leads. Having in mind a possible experimental realization, we assume that the SOI in the 
leads is weak and can be neglected. Furthermore, the magnetic field is restricted to the 
wire region. We assume that after entering the leads the electrons are independent (Fermi liquid behavior 
in higher-dimensional systems).   
For a local lead-wire coupling and in the low-energy limit, only the leads' density of states 
at the end of the leads and at the chemical potential $\mu$ (energy of incoming electrons) 
matters. For simplicity, we thus model 
the leads as semi-infinite 1D tight-binding chains ($\chi=L,R$)
\begin{eqnarray}
H_{\chi}^\mathrm{lead}&=-\tilde t_{\chi} 
\sum\limits_j\sum\limits_{\sigma}& 
\left[ d_{j+1,\sigma}^\dagger d_{j,\sigma} +\mathrm{H.c.} \right]\;,\label{Hleads}
\end{eqnarray}
with $d_{j,\sigma}^\dagger$ being the creation operator of the leads, $j=-\infty,...,0$ for 
the left lead, and $j=N+1,...,\infty$ for the right one. To prevent a proliferation of parameters 
we assume equal leads, set $\tilde t_L= \tilde t_R=t$, and measure all energies in units 
of the lead hopping $t=1$. Our energy unit is therefore of the order of 1 eV. 
At the contacts, the electrons can tunnel in and out of the wire and the Hamiltonian of 
the wire-lead coupling is given by
\begin{eqnarray}
H_\mathrm{coup}=&\sum\limits_{\sigma}& 
\big[ t_L c_{1,\sigma}^\dagger d_{0,\sigma} + t_R d_{N+1,\sigma}^\dagger 
c_{N,\sigma} +\mathrm{H.c.} \big] \;.
\label{Hdotleads}
\end{eqnarray}

The site- and bond-dependence of $\alpha_y, \alpha_z$, 
the magnetic field, and the interaction matrix elements allows us to adiabatically 
turn on and off these couplings over $m_1$ lattice sites/bonds close to the wire-lead contacts. 
This will be done in order to prevent any unwanted electron backscattering from the 
contacts and is reminiscent of the gradual confinement to the 1D geometry in heterostructures. 
We emphasize that the precise shape of the weight function with which the couplings 
are turned on and off does not have any significant effect on the 
results as long as it is sufficiently smooth. In the bulk of the wire, these parameters 
reach constant values (for details, see below).

\section{Methods}
\label{method}

\subsection{The linear conductance}

Using the Landauer-B\"uttiker approach,\cite{datta2} one can express the 
spin-dependent linear conductance $G_{\sigma,\sigma'}$ for 
{\it vanishing two-particle interaction} in terms of the transmission $\mathcal{T}_{\sigma\sigma'}(\varepsilon)$ 
\begin{equation}
G_{\sigma,\sigma'} =-\frac{\displaystyle e^2}{\displaystyle h} 
\int |\mathcal{T}_{\sigma,\sigma'}(\epsilon)|^2 f'(\epsilon) \mathrm{d}\epsilon \label{lb1}
\end{equation}
with $f(\epsilon)$ being the Fermi function. The indices $\sigma,\sigma'$ denote the $z$-component 
of the electron spin 
before entering and after leaving the quantum wire, respectively. The spin is conserved 
outside the quantum wire, since we neglect any SOI in the leads as well as spin relaxation.
Using the Feshbach projection,\cite{enss1} it is easy to show that the transmission is connected 
to the $(1,N)$ matrix element of the retarded single-particle Green's 
function $\mathcal{G}^{\sigma,\sigma'}(\varepsilon+i0)$ of the entire system (including the leads)
\begin{equation}
\mathcal{T}_{\sigma,\sigma'}(\varepsilon)=2t_L t_R \sin(k_\varepsilon)
\mathcal{G}^{\sigma,\sigma'}_{1,N}(\varepsilon+i0) \label{trans}
\end{equation}
with $k_\varepsilon=\arccos(-\varepsilon/2)$.

At $T=0$, on which we focus from now on, the derivative of the Fermi function is 
a $\delta$-function and Eq.~(\ref{lb1}) 
simplifies to
\begin{equation}
G_{\sigma,\sigma'} = \frac{\displaystyle e^2}{\displaystyle h} \left| \mathcal{T}_{\sigma,\sigma'}(\mu) 
\right|^2 \; , 
\label{lb2}
\end{equation}
with the chemical potential given by the Fermi momentum, $\mu=-2\cos k_F$.
For noninteracting leads, this relation holds even 
if the two-particle interaction in the wire is finite,\cite{Oguri} where $\mathcal{G}$ in Eq.~(\ref{trans}) is 
the {\it interacting} Green's function of the entire system. To compute the latter we first integrate 
out the leads by projection (see the next subsection) and treat the remaining interacting system of size $N$ using 
the approximate fRG procedure.   

Using standard Feshbach projection, the effect of the leads can be cast in energy dependent 
contributions to the $(1,1)$ and $(N,N)$  matrix elements of the self-energy.\cite{enss1} 
They read
\begin{eqnarray}
\left(\Sigma_{\rm lead}\right)_{1,1}^{\sigma,\sigma'}(z) & = & \delta_{\sigma,\sigma'} 
t_L^2 g(z) \; , \nonumber \\
\left(\Sigma_{\rm lead}\right)_{N,N}^{\sigma,\sigma'}(z) & = & \delta_{\sigma,\sigma'} 
t_R^2  g(z) 
\label{SEleads}
\end{eqnarray}
with the Green's function 
\begin{eqnarray}
g (z) = \displaystyle z+\mu \mp \sqrt{(z+\mu)^2 - 4} 
\end{eqnarray}
of the semi-infinite leads taken at the first lattice site. The sign must be chosen such that 
$\lim\limits_{|z|\rightarrow\infty} g_{\sigma,\sigma}(z)=0$. 
In the following, we absorb the contribution of the leads to the self-energy into the 
noninteracting propagator $\mathcal{G}_0$ of the wire. The full 
interacting Green's function is given by the Dyson equation
\begin{equation}
\mathcal{G}=\left(\mathcal{G}_0^{-1}-\Sigma\right)^{-1}\; .\label{dyson}
\end{equation}

\subsection{Functional renormalization group} 
\label{fRG}

We briefly describe the fRG method used here\cite{footfoot} to approximately 
compute $\mathcal{G}$.\cite{SalmhoferHonerkamp,fRG-skript} 
Detailed accounts of the application of this method to inhomogeneous,
interacting quantum wires have been given in the last few 
years.\cite{enss,andergassen1} Including the SOI and the magnetic field does only require 
minor extensions.

Starting point of the fRG scheme is the propagator 
$\mathcal{G}_0$ which follows from the noninteracting Hamiltonian Eq.~(\ref{H0}) and 
the self-energy contribution of the leads Eq.~(\ref{SEleads}). It is supplemented with a 
cutoff $\Lambda$ such that all modes with Matsubara frequencies $|\omega|<\Lambda$ 
are suppressed, i.e.
\begin{equation}
\mathcal{G}_0^\Lambda(i\omega)=\Theta(|\omega|-\Lambda)\mathcal{G}_0(i\omega)\;,
\end{equation}
where $\Lambda$ runs from $\infty$ to $0$. 
Inserting $\mathcal{G}_0^\Lambda$ in the generating functional of the one-particle irreducible vertex 
functions, one obtains an infinite hierarchy of coupled differential equations 
for the vertex functions by differentiating the generating functional with respect to $\Lambda$ 
and expanding it in powers of the external fields. To obtain a manageable set of flow equations, this hierarchy 
must be truncated. In a first step, we neglect the three-particle vertex $\Gamma_3^\Lambda$, 
since it is zero at $\Lambda=\infty$ and is generated only from terms of third order in 
the two-particle vertex $\Gamma^\Lambda$, which are small as long as $\Gamma^\Lambda$ does not become 
too large. 

For arbitrary local $U_1$, nearest-neighbor interaction $U_2$, and filling $n\in (0,2)$ of the band, 
the flow of the two-particle vertex must be kept at least to lowest (that is second) order 
to correctly obtain the scaling behavior (for large $N$) of correlation 
functions to leading order in the interaction. In the absence of SOI 
this is already known from the so-called ``g-ology'' model.\cite{Solyom} An RG analysis of 
this model shows that the two-particle scattering with momentum transfer $2 k_F$ of electrons 
with opposite spin, the so-called $g_{1,\perp}$ term, flows to zero. If one is interested in 
correlation functions of large systems, this scaling must be captured by any 
sensible approximation. For the extended Hubbard model and vanishing SOI, this has been done in 
Ref.~\onlinecite{andergassen1} using fRG. As the parameterization of the two-particle vertex used 
there relies on spin conservation, it cannot easily be extended to the present situation with SOI. 
Here, we are thus forced to proceed differently. 

In the extended Hubbard model the coupling $g_{1,\perp}$ is given by  
\begin{eqnarray}
g_{1,\perp} = U_1 + 2 U_2 \cos{(2 k_F)} \; .
\label{g_1def}
\end{eqnarray}
If it is zero initially, it will not get generated in a lowest-order RG analysis of the corresponding 
``g-ology'' model (for vanishing SOI). If one is interested in the behavior 
of correlation functions to leading order in the interaction, the flow of the vertex can then 
be neglected altogether. A vanishing $g_{1,\perp}$ is achieved, if we stick to parameters 
$U_1$, $U_2$, and $k_F$ 
such that the right hand side of Eq.~(\ref{g_1def}) is zero. Neglecting terms of order $U^2 \alpha$, 
where $U$ stands for either $U_1$ or $U_2$ and $\alpha$ for either $\alpha_z$ or $\alpha_y$ 
the same reasoning holds if the SOI is included. When studying the effect of the two-particle 
interaction for large systems (Luttinger liquid behavior in the scaling limit), we thus 
exclusively consider the case 
\begin{eqnarray}
\label{condi}
U_2=-U_1/[2 \cos{(2 k_F)}] 
\end{eqnarray}
and neglect the 
flow of the two-particle vertex. 
This assumption does not affect the asymptotic power-law scaling of the conductance 
for vanishing SOI.\cite{andergassen1} We expect the same to hold for nonzero SOI and thus 
focussing on this situation does not present a severe constraint for our purposes.  
An fRG analysis of the flow of the two-particle vertex in the 
presence of SOI which would allow to investigate novel phases resulting from the interplay of 
the two-particle interaction and the SOI\cite{mikhail} is left for further studies. 

For wires of the 
order of a hundred lattice sites, the flow of the components of the two-particle vertex is cut off 
on fairly large energy scales ($\propto 1/N$), affects correlation functions (such as the single-particle 
Green's function we are aiming at) only weakly, and can thus be neglected even if one chooses parameters such that 
Eq.~(\ref{condi}) does not hold. Our approximation contains at least all terms 
of first order perturbation theory in the two-particle interaction.\cite{enss,andergassen1}     

Within these approximations the self-energy (negative of the one-particle vertex) 
becomes frequency independent\cite{enss,andergassen1}, and its flow equation reads  
\begin{equation}
\frac{\partial}{\partial\Lambda}\Sigma_{1',1}^\Lambda=-\frac{1}{2\pi}\sum\limits_{\omega=\pm\Lambda}\sum\limits_{2,2'}e^{i\omega 0^+}{\mathcal{G}}_{2,2'}^\Lambda(i\omega)\Gamma_{1',2';1,2} \;, \label{DGLsigma}
\end{equation} 
where the indices $1,1',2,2'$ label the quantum numbers ${j,\sigma}$ and 
\begin{equation}
{\mathcal{G}}^\Lambda(i\omega)=\left[\mathcal{G}_0^{-1}-\Sigma^\Lambda\right]^{-1} \; .
\end{equation}
One starts the numerical integration at a large initial value $\Lambda_0\sim 10^8$ and 
integrates down to $\Lambda=0$. Following the description in Ref.~\onlinecite{enss}, 
we add a one-particle potential $\nu$ to the Hamiltonian $H_1$ and $H_2$, such that the starting
value of the self-energy, which accounts for the finite contribution resulting from 
the integration of Eq.~(\ref{DGLsigma}) over the interval $(\infty,\Lambda_0]$, vanish. 
The initial condition for $\Lambda_0 \to \infty$ then reads
\begin{eqnarray}
\Sigma_{1,1'}^{\Lambda_0} = 0 \; .
\end{eqnarray}
The self-energy $\Sigma^{\Lambda=0}$ at the end of the flow can be considered as an approximation
to the full self-energy. Using the Dyson equation (\ref{dyson}), the Green's function entering 
the expression for the $T=0$ linear conductance can be computed. 

For our model with only local 
and nearest neighbor interactions the self-energy is a tridiagonal matrix in the Wannier basis 
states, with each entry being a $2 \times 2$ matrix in spin space. As $\Sigma$ is frequency independent 
within our approximation, its matrix elements can be interpreted as the interaction induced renormalizations 
of the parameters of the noninteracting model. Depending on the matrix element considered these are the 
SOI and spin conserving hoppings, magnetic fields as well as on-site 
potentials all being position dependent.\cite{enss, andergassen1} This implies that 
to approximately obtain the $T=0$ conductance of 
an interacting system  the Green's function of an effective noninteracting problem with renormalized 
parameters must be computed. This is similar to the Hartree-Fock approximation. However, we emphasize 
that in our approximate fRG a different class of diagrams is resummed, this way avoiding the artifacts 
known to emerge if the Hartree-Fock approximation is used for problems of interacting 
electrons in low dimensions.       

For half filling ($n=1$), the electron correlations drive our system towards a Mott insulator state (details 
depend on the parameters $U_1$ and $U_2$), which becomes 
of relevance for sufficiently large $N$.\cite{KS} As we are here interested in spin dependent transport through 
{\it metallic} wires we only consider fillings away from $n=1$ when studying large $N$.\cite{footy}   

\section{Results}
\label{results}

Following the steps described in the last section, we calculate the spin resolved 
linear conductance $G_{\sigma,\sigma'}$ with $\sigma,\sigma = {\uparrow,\downarrow}$ (with respect to the 
$z$-direction), as a function of the 
system's chemical potential $\mu$ (energy of incoming electrons) and the system size $N$.
The total conductance is given by the sum of the four components
\begin{equation} G_\mathrm{total}=G_{\uparrow\uparrow}+G_{\uparrow\downarrow}+G_{\downarrow\uparrow}
+G_{\downarrow\downarrow}\;. \label{Gtotal}
\end{equation}
As our leads are free of SOI, the definition of the spin polarization does not involve the spin expectation 
value. This has to be contrasted to the definition of Ref.~\onlinecite{streda} for a system without leads.
Due to the choice of the $z$-axis as the spin quantization axis, the spin polarization in $z$-direction 
can be defined as the ``normalized'' difference between the probability that an electron enters the right lead 
with spin up and the probability that it enters with spin down 
\begin{equation}
P_z=\frac{\displaystyle G_{\uparrow\uparrow}+G_{\downarrow\uparrow}-G_{\uparrow\downarrow}-
G_{\downarrow\downarrow}}{\displaystyle G_\mathrm{total}}\;. \label{polar}
\end{equation}
The conductance components and spin polarization in the $x$- and $y$-direction can be obtained by a simple base 
transformation. The transmissions which need to be inserted into Eq.~(\ref{lb2}) are 
\begin{eqnarray}
\mathcal{T}_{\uparrow\uparrow}^{(x)}&=&\left(\mathcal{T}_{\uparrow\uparrow}+\mathcal{T}_{\uparrow\downarrow}+\mathcal{T}_{\downarrow\uparrow} +\mathcal{T}_{\downarrow\downarrow} \right)/2 \; ,\nonumber\\
\mathcal{T}_{\uparrow\downarrow}^{(x)}&=&\left(\mathcal{T}_{\uparrow\uparrow}-\mathcal{T}_{\uparrow\downarrow}+\mathcal{T}_{\downarrow\uparrow} -\mathcal{T}_{\downarrow\downarrow} \right)/2 \; .\nonumber\\
\mathcal{T}_{\downarrow\uparrow}^{(x)}&=&\left(\mathcal{T}_{\uparrow\uparrow}+\mathcal{T}_{\uparrow\downarrow}-\mathcal{T}_{\downarrow\uparrow} -\mathcal{T}_{\downarrow\downarrow} \right)/2 \; ,\nonumber\\
\mathcal{T}_{\downarrow\downarrow}^{(x)}&=&\left(\mathcal{T}_{\uparrow\uparrow}-\mathcal{T}_{\uparrow\downarrow}-\mathcal{T}_{\downarrow\uparrow} +\mathcal{T}_{\downarrow\downarrow} \right)/2
\end{eqnarray}
and
\begin{eqnarray}
\mathcal{T}_{\uparrow\uparrow}^{(y)}&=&\left(\mathcal{T}_{\uparrow\uparrow}-i\mathcal{T}_{\uparrow\downarrow}+i\mathcal{T}_{\downarrow\uparrow} +\mathcal{T}_{\downarrow\downarrow} \right)/2 \; ,\nonumber\\
\mathcal{T}_{\uparrow\downarrow}^{(y)}&=&\left(-i\mathcal{T}_{\uparrow\uparrow}+\mathcal{T}_{\uparrow\downarrow}+\mathcal{T}_{\downarrow\uparrow} +i\mathcal{T}_{\downarrow\downarrow} \right)/2 \; , \nonumber\\
\mathcal{T}_{\downarrow\uparrow}^{(y)}&=&\left(i\mathcal{T}_{\uparrow\uparrow}+\mathcal{T}_{\uparrow\downarrow}+\mathcal{T}_{\downarrow\uparrow} -i\mathcal{T}_{\downarrow\downarrow} \right)/2 \; , \nonumber\\
\mathcal{T}_{\downarrow\downarrow}^{(y)}&=&\left(\mathcal{T}_{\uparrow\uparrow}+i\mathcal{T}_{\uparrow\downarrow}-i\mathcal{T}_{\downarrow\uparrow} +\mathcal{T}_{\downarrow\downarrow} \right)/2 \; ,
\end{eqnarray}
where $\mathcal{T}_{\sigma,\sigma'}$ denotes the transmission with respect to the $z$-direction. The corresponding 
polarizations follow as in Eq.~(\ref{polar}) with $G$ replaced by $G^{(x)}$ and $G^{(y)}$, respectively.

\subsection{Vanishing Coulomb interaction}

\subsubsection{A potential step}

\begin{figure}[tb]
\includegraphics[width=0.45\textwidth,clip]{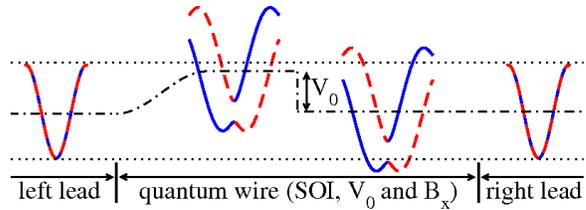}
\caption{(Color online) Sketch of the system with SOI and parallel magnetic field in the wire.
The leads have a cosine-like dispersion, whereas the ``local'' dispersion in the quantum wire has two 
nondegenerate branches [$s=+$ (solid line), $s=-$ (dashed line)].
A potential step which is turned on smoothly at the left contact and turned of sharply in the middle of the wire 
is indicated by the dashed-dotted line. The energies of the incoming and outgoing electrons are confined 
within the dotted lines.}
\label{system_leads}
\end{figure}

To investigate whether spin polarized currents can be achieved using a mechanism similar to that 
of Ref.~\onlinecite{streda} in the presence of nonmagnetic leads, we consider a setup that is 
close in spirit to the one studied there. We therefore add a potential step to the wire which can 
be used to shift the energy into the region of only doubly-degenerate states as sketched in  
Fig.~\ref{system_leads}. To prevent any backscattering at the left contact it is turned on smoothly 
there and turned off sharply in the middle of the wire  (dashed-dotted line in Fig.~\ref{system_leads}). 
Due to the smooth  variation of the system parameters at the left contact, the cosine-like dispersion of 
the leads is ``adiabatically'' transformed into the ``local'' dispersion  in the wire. A similar ``adiabatic'' 
transition occurs at the right contact. As the achieved wire dispersion is an essential ingredient 
of the spin filter to perform properly,\cite{streda,birkholz1} modelling a smooth variation at the contacts 
is mandatory. However, assuming a gradual crossover from higher-dimensional leads to the
1D wire appears to be quite natural in heterostructures.   

\begin{figure}[tb]
\includegraphics[width=0.4\textwidth,clip]{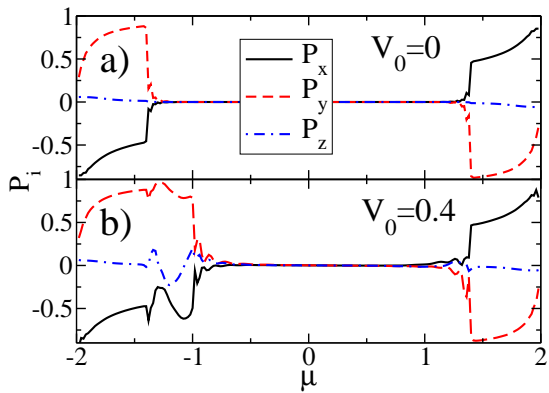}
\caption{(Color online) Spin polarization in $x$-, $y$- and $z$-direction as a function of $\mu$ for a 
homogeneous system [$V_0=0$, a)] and a system 
with potential step [$V_0=0.4$, b)] for system parameters $N=100,m_1=20, t_L=t_R=1,\alpha_y=0,\alpha_z=0.8$, and $\gamma B_x=0.6$. 
Inside the region with two conducting channels, there is no significant spin polarization. The high and low energy 
polarization is independent from $V_0$ in good approximation. The potential leads 
to effects for $\mu\in [-2+\gamma B_x,-2+\gamma B_x+V_0]$.}
\label{pol_Bx_SOI_V0}
\end{figure}

Fig.~\ref{pol_Bx_SOI_V0} shows the spin 
polarization $P_i$, $i=x,y,z$, for a homogeneous system 
[$V_0=0$,  Fig.~\ref{pol_Bx_SOI_V0}a)] and a 
system with potential step [$V_0=0.4$,  Fig.~\ref{pol_Bx_SOI_V0}b)] for system parameters 
$N=100,m_1=20, \alpha_y=0,\alpha_z=0.8$, and $\gamma B_x=0.6$. 
We chose the value for $\alpha_z$ to be larger than the spin-orbit parameter $\alpha\sim 0.1\,\mathrm{eV}$ 
extracted from Ref.~\onlinecite{grundler} for bulk InAs, 
since the SOI is significantly increased in semiconductor heterostructures.
The lead-wire tunnel contacts modelled by $t_{L/R}$ are assumed to be ``perfect''
$ t_L=t_R=1$, a situation on which we focus from now on. 
The curves show a large spin polarization in $x$- and $y$-direction for 
$\mu\not\in [-2+\gamma B_x+V_0, 2-\gamma B_x]$, i.e.~as long as there is only one conducting 
channel. This polarization is a ``trivial'' band effect due to the Zeeman splitting. Without SOI, 
one obtains a perfect spin-polarization in $x$-direction, $P_x=\pm 1$, in this interval 
[see Fig.~\ref{pol_Bx_SOI_V1}a)]. 
For finite SOI, the spin is rotated out of the $x$-direction leading to non-vanishing components 
$P_y$ and $P_z$. Similarly to the continuum situation, \cite{birkholz1} $\alpha_z$ ($\alpha_y$) mainly 
leads to a spin rotation into the $y$- ($z$-) direction.

The most interesting energy regime in connection with Ref.~\onlinecite{streda} is 
$\mu\in [-2+\gamma B_x,-2+\gamma B_x+V_0]$ in Fig.~\ref{pol_Bx_SOI_V0}b). In this interval 
of width $V_0$, we observe nonvanishing, but strongly oscillating components of the polarization. 
The oscillations can be traced back to those of the conductance components and are absent in the 
continuum model of Ref.\ \onlinecite{streda}. By a mechanism similar 
to the one discussed in Ref.~\onlinecite{streda}, we obtain a nonvanishing 
polarization. As was pointed out in Ref.~\onlinecite{birkholz1} for the lead-less continuum model, the 
polarization $P_x$ (in $x$-direction) only depends on the absolute value $\alpha$ and not on the 
direction of the effective SOI field and $|P_y/P_z|=|\alpha_z/\alpha_y|$. Averaging over the polarization 
oscillations, this holds in very good approximation also for the lattice model with leads, although the 
polarization here is defined 
by Eq.~(\ref{polar}) and not in terms of spin expectation values.\cite{streda} 
Because of the large oscillations (absent in the lead-less continuum model) the spin 
polarization reacts very sensitive to changes in the chemical potential $\mu$. This could be 
of interest to control the spin polarization in future spintronic devices.

\subsubsection{A single impurity}
\label{jojo}

Another interesting polarization effect occurs, if we insert a localized impurity of strength $V_1$ in 
our system with SOI and parallel magnetic field. The exact position of the impurity within the bulk part 
of the wire does not matter and we here locate it in the middle of the wire.

\begin{figure}[tb]
\includegraphics[width=0.4\textwidth,clip]{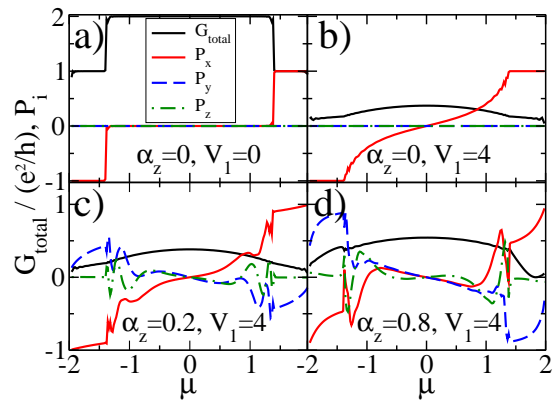}
\caption{(Color online) Conductance and spin polarization for a system of $N=101$ lattice sites ($m_1=20$) with a single impurity 
$V_1$ in the middle of the system in presence of a magnetic field $\gamma B_x=0.6$ and different SOI couplings 
$\alpha_z$ ($\alpha_y=0$). a) The homogeneous system shows a perfect step-like polarization in $x$-direction. 
b) The shape of $P_x$ is washed out in presence of a strong single impurity $V_1=4$ and the conductance is severely 
decreased. c) Small SOI coupling $\alpha_z=0.2$ leads to finite $P_y$, $P_z$ and polarization oscillations. 
d) The oscillations are enhanced by increasing $\alpha_z$.}
\label{pol_Bx_SOI_V1}
\end{figure}

Fig.~\ref{pol_Bx_SOI_V1} shows the conductance and spin polarization of a system with $N=101$ 
lattice sites ($m_1=20$) with a single impurity $V_1$ for a magnetic field 
$\gamma B_x=0.6$ and different SOI couplings $\alpha_z$ ($\alpha_y=0$). 
The system without SOI and without any impurity ($V_1=0$) 
shows a perfect step-like polarization in $x$-direction (``trivial'' band effect), whereas $P_y=P_z=0$ 
[see Fig.~\ref{pol_Bx_SOI_V1}a)]. The step is smeared out and $P_x$ can be tuned smoothly
in presence of an intermediate to strong single impurity $V_1=4$ while the total conductance is severely decreased in  this 
case [see Fig.~\ref{pol_Bx_SOI_V1}b)].
For small SOI, $\alpha_z=0.2$, $P_x$ is still the dominant polarization component, 
but $P_y$ and $P_z$ become finite. 
All polarization components show oscillations for $\mu\in [-2+\gamma B_x,2-\gamma B_x]$, which 
become heavily pronounced at the edges of this interval [see Fig.~\ref{pol_Bx_SOI_V1}c)]. 
For large SOI [see Fig.~\ref{pol_Bx_SOI_V1}d)], $\alpha_z=0.8$, we observe the same behavior 
for $\mu\notin [-2+\gamma B_x,2-\gamma B_x ]$ as in the impurity-free case [see Fig.~\ref{pol_Bx_SOI_V0}a)] 
with $P_z$ playing only a minor role. The oscillations of the polarization components are more pronounced 
compared to the case with small $\alpha_z$, especially for $P_x$. Moreover, 
the total conductance is enlarged due to the smaller ratio of $V_1$ and the effective hopping 
$\sqrt{t^2+\alpha_z^2}$. Due to the large oscillations, each spin polarization component reacts 
very sensitive to changes in $\mu$ and can therefore be tuned by adjusting $\mu$.

In this section we studied systems with $N \sim {\mathcal O}(10^2)$ lattice sites, corresponding to wires of the order 
of tens of nanometers. For vanishing two-particle interaction increasing the number of lattice sites does 
not affect the results obtained here.

\subsection{The effect of the Coulomb interaction}

We now add the terms $H_1$ and $H_2$ given in Eqs.~(\ref{H1}) and (\ref{H2}) to our Hamiltonian and use the fRG  
to approximately compute the conductance. We first study short quantum wires with $N \sim {\mathcal O}(10^2)$ 
lattice sites and investigate how the energy regime (of the incoming electrons) in which spin polarized currents can
be obtained is modified by the Coulomb interaction. Changes can be traced back to the interaction induced 
renormalization of the parameters of the noninteracting model.\cite{JensDr} 

In a second step we focus on a fixed chemical potential, at which 
spin polarization is observed and study how this is modified if the system size $N$ is increased. The 
inverse of $N$, more precisely $\delta \propto v_F/N$ (with the Fermi velocity $v_F$), presents an infrared 
energy scale in our setup. In the absence of SOI and a magnetic field, inhomogeneities, such as single impurities 
and potential steps, are known to lead to a power-law suppression of the conductance as a function of 
infrared energy scales.\cite{KaneFisher,Furusaki,enss,andergassen1} We demonstrate that while the total conductance 
shows a similar behavior for finite SOI, the polarization does not display scaling behavior.       

To model a gradual transition from a higher-dimensional to a 1D system at the lead-wire contacts, 
we gradually increase $U_1$ and $U_2$ over $m_1$ lattice sites starting at the contacts.
This prevents electron backscattering at the contacts due to the inhomogeneous two-particle 
interaction and one achieves unitary total conductance in the absence of an external single-particle potential.  
As the details of the variation of the interaction do not matter as long as it is sufficiently smooth,\cite{Katharina} we chose 
the same weight function as for the single-particle parameters. For an increasing chain length, the number of 
lattice sites over which the two-particle interaction and the single-particle parameters are turned on and off 
close to the contacts must be increased.\cite{Katharina}

\subsubsection{Short wires}

\begin{figure}[tb]
\begin{center}
\includegraphics[width=0.4\textwidth,clip]{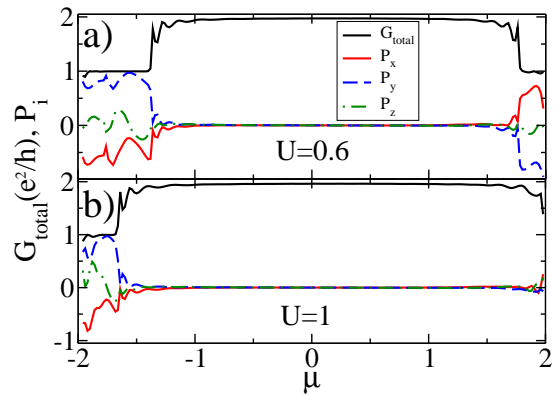}
\end{center}
\caption{(Color online) Total conductance and spin polarization as a function of the chemical potential $\mu$ for 
a system with $N=100$ lattice sites ($m_1=20$), potential step $V_0=0.4$, magnetic field $\gamma B_x=0.6$ and different Coulomb 
interaction $U=U_1=2U_2$. Increasing the Coulomb interaction [from a) to b)] for constant SOI 
$\alpha_z=0.8$ ($\alpha_y=0$) leads to a broadening of the energy regime with two open conducting channels. 
The spin polarization vanishes in this regime, but reveals a strong dependence on $\mu$ in the regime with only 
one conducting channel.}
\label{pol_V0_U}
\end{figure}

Fig.~\ref{pol_V0_U} shows the total conductance and spin polarization as a function of the chemical 
potential $\mu$ for a system with the same parameters as in Fig.~\ref{pol_Bx_SOI_V0}b) but for 
nonvanishing Coulomb interaction. 
We consider a constant 
ratio of local and nearest-neighbor interaction, $U_1/U_2=2$ (which appears to be a rather physical value), 
denote the local interaction $U_1$ by $U$ 
and study its affect on $G_\mathrm{total}$ and $P_i$, $i=x,y,z$. Comparing Figs.~ \ref{pol_Bx_SOI_V0}b), \ref{pol_V0_U}a) 
and \ref{pol_V0_U}b), one notes that an increase of $U$ leads to a broadening of the energy regime with two 
open conducting channels, that is ``perfect'' conductance $G_\mathrm{total} \approx 2\,e^2/h$, and vanishing spin polarization. 
The Coulomb interaction thus leads to a decrease of the energy range in which spin polarization can be achieved. 
We emphasize, that within our approximation the above behavior can be understood in terms of transport 
through a quantum wire with {\it vanishing} two-particle interaction, but renormalized 
single-particle parameters (regular and spin-flip hoppings, magnetic field, and on-site potentials all of 
them depending on the lattice site index $j$) given by the self-energy at the end of the fRG flow.\cite{JensDr} 

\begin{figure}[tb]
\includegraphics[width=0.4\textwidth,clip]{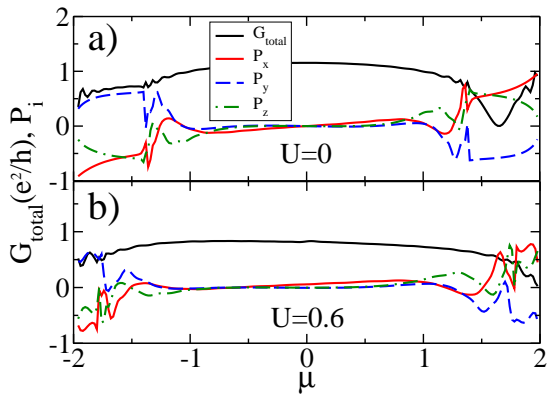}
\caption{(Color online) Total conductance and spin polarization components for a system with $N=101$ lattice 
sites ($m_1=20$), SOI coupling parameters $\alpha_y=\alpha_z=0.5$, magnetic field in $x$-direction, $\gamma B_x=0.6$, 
and a single impurity $V_1=2$ in the middle of the system. Increasing [from a) to b)] the two-particle 
interaction $U$ leads to an increase of the energy regime with two conduction channels, but an overall 
decrease of the total conductance.} 
\label{pol_U1_Bx_SOI_V1}
\end{figure}

We now investigate the influence of the Coulomb interaction on the transport properties of a short wire with 
a single impurity $V_1$ in the middle of the quantum wire. Fig.~\ref{pol_U1_Bx_SOI_V1} shows the total 
conductance $G_\mathrm{total}$ and polarization components $P_i$, $i=x,y,z$, for a system with $N=101$ 
lattice sites ($m_1=20$), SOI parameters $\alpha_y=\alpha_z=0.5$ and magnetic field $\gamma B_x=0.6$ in $x$-direction 
for $V_1=2$. Increasing the Coulomb interaction, we observe two tendencies. The first one is the increase of 
the energy regime with two conduction channels already found in the presence of the potential step. As the 
electrons scatter off the single impurity this does not lead to unitary conductance in the present case 
(in contrast to the case of a potential step). The second effect is an overall decrease of the total conductance 
with increasing $U$. This was to be expected as it is known that 
the effective strength of inhomogeneities increases in the presence of Coulomb 
correlations, eventually leading to power-law suppression for sufficiently large $N$ 
(that is sufficiently small energies $\delta$).\cite{KaneFisher,Furusaki,enss,andergassen1}   

For the fairly short wires studied in this subsection, results from first order perturbation theory 
in the two-particle interaction (for the self-energy) lead to qualitatively similar results as those obtained 
by our approximate fRG scheme. This does no longer hold for the longer wires studied next (Luttinger liquid 
behavior).

\subsubsection{Luttinger liquid behavior in long wires}

In the absence of SOI and a magnetic field, the linear conductance of a Luttinger liquid 
wire with a single impurity $V_1$ is known to show  power-law suppression 
as a function of an infrared energy scale (e.g.~the temperature or $\delta \propto v_F/N$), provided the 
latter is sufficiently small (scaling regime).\cite{KaneFisher,Furusaki,enss,andergassen1} 
Surprisingly $G_{\rm total}$ vanishes in the asymptotic low-energy limit even for small $V_1$. However, the energy 
scale beyond which scaling holds depends on $V_1$ and becomes exponentially small for small $V_1$, implying that 
exponentially long chains must be studied to observe power-law behavior. The 
scaling exponent is a function of the interaction and the filling, but is independent of the strength of the bare 
impurity.\cite{KaneFisher,Furusaki,enss,andergassen1} For the extended Hubbard model 
and our choice of the ratio $U_1/U_2=U/U_2=-2 \cos{(2 k_F)}$ (vanishing two-particle backscattering $g_{1\perp}$; see Sect.~\ref{fRG}) 
it is given by 
\begin{equation}
2 \alpha_B=-\frac{\mu^2-4\cos(\pi n)}{(2-\mu^2)(2\pi t\sin(\pi n/2))} \; U \label{beta}
\end{equation} 
to leading order in $U$.\cite{andergassen1} It was shown earlier that the approximate fRG procedure captures this power-law 
behavior and correctly reproduces the scaling exponent to order $U$.\cite{enss,andergassen1}  
In order to be able to compare our results to Eq.~(\ref{beta}) we need to know $n$ and 
therefore tune the additional one-particle potential $\nu(U,\mu)$, introduced in Sect.~\ref{fRG}, such 
that the filling of the 1D quantum wire with electrons in presence of $U_1$ and $U_2$ corresponds to the 
filling $n=2 \arccos{(-\mu/2)}/\pi$ of the noninteracting leads at given $\mu$. 
Following Ref.~\onlinecite{andergassen1}, the starting values of the self-energy 
at lattice site $j$ due to integration of the flow equations from $\infty$ down to 
$\Lambda_0$ are given by
\begin{eqnarray}
\Sigma^{\sigma\sigma';\Lambda_0}_{j,j'} & = &
\left(\frac{\displaystyle 1}{\displaystyle 2}U_{1,j}+2U_{2,j}\right) \delta_{j,j'} 
\delta_{\sigma,\sigma'},\; j\in \{ 2,\ldots,N-1 \} \; , \nonumber\\
\Sigma^{\sigma\sigma';\Lambda_0}_{j,j'} & = & \left(\frac{\displaystyle 1}{\displaystyle 2}U_{1,j}+U_{2,j}\right) 
\delta_{j,j'}\delta_{\sigma,\sigma'},\; j =1,N\; ,
\end{eqnarray}
up to corrections of order $1/\Lambda_0$.

\begin{figure}[tb]
\includegraphics[width=0.4\textwidth,clip]{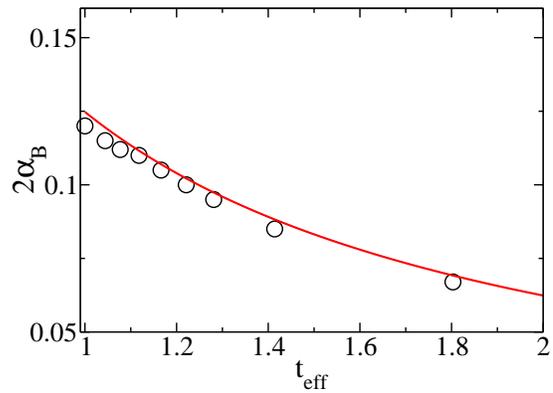}
\caption{(Color online) 
Dependence of the scaling exponent on the strength of the SOI (varying $\alpha_y$ as well as 
$\alpha_z$) for $U=0.3$ and filling $n=0.75$. The circles are the data extracted from fitting 
$G_{\rm total}(\delta)$ by a power law for systems of a few thousand lattice sites. The error 
is of the order of the symbol size.  
The effective hopping $t_\mathrm{eff}$ depends on $\alpha_y$ as well as $\alpha_z$ and 
is defined in Eq.~(\ref{teff}).  The line shows the analytic expression 
Eq.~(\ref{beta}) with $t \to t_\mathrm{eff}$. The systematic deviation of the fRG data from 
the analytic expression can be explained by 
higher order corrections in $U$ included in the numerical data, but not in Eq.~(\ref{beta})}.
\label{powerlaw}
\end{figure} 

\begin{figure}[tb]
\includegraphics[width=0.4\textwidth,clip]{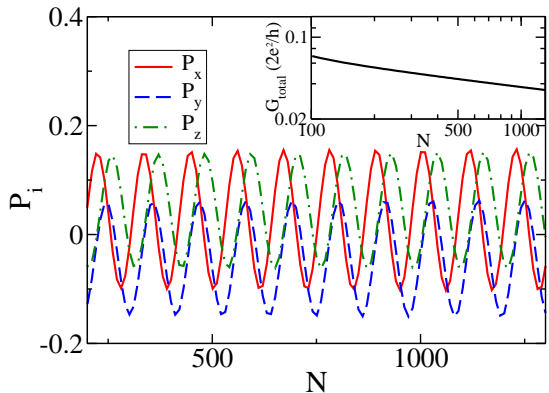}
\caption{(Color online) 
Main plot: Dependence of the polarization on the system size $N$ for 
for $V_1=8$, $\alpha_y = \alpha_z=0.5$, $\gamma B_x=0.6$, $n=0.9$, and $U=0.8$.
Inset: The total conductance as a function of $N$ on a log-log scale. The parameters 
are as in the main plot.}  
\label{pollarge}
\end{figure}

We first investigate if power-law scaling in the presence of a single impurity is also found for 
nonvanishing SOI, but zero magnetic field, a situation in which no spin polarization is found. 
To this end we compute the total conductance as a function of $\delta \propto v_F/N$ for fixed (small) 
$U \lessapprox 1$ (range of applicability of our approximate fRG procedure), fixed filling $n \neq 1$ 
(see the discussion in Sect.~\ref{fRG}), fixed SOI, and fixed intermediate 
to large $V_1$ [such that the scaling regime is reached for $N \sim {\mathcal O}(10^3)$]. 
The data for $G_{\rm total}(\delta)$ can be fitted by a power-law and we extract the asymptotic 
exponent. An example of the power-law behavior (as a function of $N \sim \delta^{-1}$) 
in the case of an additional magnetic field is  shown in the inset of  Fig.~\ref{pollarge}.  
The scaling exponent depends on the SOI via an effective renormalized spin conserving hopping    
\begin{equation}
t_\mathrm{eff}=\sqrt{\alpha_y^2+\alpha_z^2+t^2}\;, \label{teff}
\end{equation}
that is, the dependence of $2 \alpha_B$ is given by Eq.~(\ref{beta}) with $t$ replaced by 
$t_\mathrm{eff}$. This is shown in Fig.~\ref{powerlaw} for $U=0.3$, $n=0.75$, and different $\alpha_y$ 
as well as  $\alpha_z$. The systematic deviation of the data from the analytic expression 
(\ref{beta}) with $t \to t_\mathrm{eff}$ can be explained by higher order corrections in $U$ 
included in the numerical data, but not in Eq.~(\ref{beta}). 
The error of the fRG exponents extracted from fitting $G_{\rm total}(\delta)$ for 
systems of a few thousand lattice sites is of the order of the symbol size. 
 
\begin{figure}[tb]
\includegraphics[width=0.365\textwidth,clip]{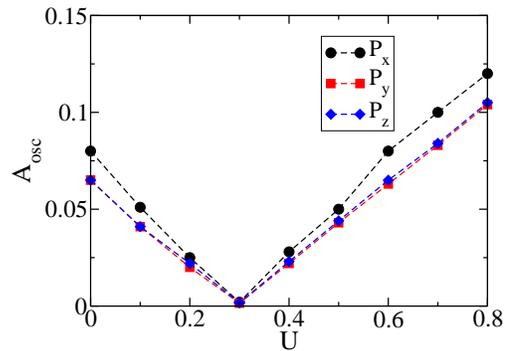}
\caption{(Color online) Amplitudes $A_\mathrm{osc}$ of the oscillations of the three components of the polarization 
(see main part of Fig.~\ref{pollarge}) as a function of the two-particle interaction $U$ for the same parameters as in 
Fig.~\ref{pollarge}. Because of $\alpha_y = \alpha_z$, the amplitudes of the polarization in $y$- and $z$-direction 
are supposed to be the same. The discrepancy of these two data sets thus gives an estimate how accurately the 
amplitude can be read off from the numerical data.}  
\label{amplifig}
\end{figure}

Next we consider the case of a single impurity in the presence of SOI and a magnetic field in 
$x$-direction, a situation with nonvanishing polarization (see the discussion in Sec.~\ref{jojo}). The main part 
of Fig.~\ref{pollarge} shows 
the $N \sim \delta^{-1}$ dependence of the three components of the polarization for $V_1=8$, 
$\alpha_y = \alpha_z=0.5$, $\gamma B_x=0.6$, $n=0.9$, and $U=0.8$. In the inset we present the corresponding 
total conductance on a log-log-scale. While the latter shows a clear indication of power-law 
suppression, the polarization oscillates with a constant amplitude $A_\mathrm{osc}$. 
A similar oscillation is found for $U=0$. 
In fact, the amplitude of the oscillation shows a nonmonotonic dependence 
on $U$. Starting from $U=0$, it first decreases linearly with increasing $U$ up to $U=0.3$, but increases for 
$U>0.3$ (see Fig.~\ref{amplifig}). 
For $U=0.8$ as shown 
in Fig.~\ref{pollarge} the amplitude is roughly a factor of 1.5 larger than in the noninteracting case. 
Fig.~\ref{pol_U1_Bx_SOI_V1} indicates that the details of this behavior depend on the filling 
(chemical potential of the incoming electrons).
The above result implies that 
although the total current through (total linear conductance of) a quantum wire with a single impurity, SOI, 
and a magnetic field in the direction of the wire is generically strongly reduced as a function 
of $N$ in the presence of the Coulomb interaction, the degree of spin polarization of the current stays 
constant. We note that the increase of  $A_\mathrm{osc}$ as a function of $U$ (see Fig.\ \ref{amplifig}) 
has to be considered with 
caution as our approximation is only valid for sufficiently small $U$, while the increased
polarization (compared to the noninteracting one) requires a small to intermediate $U$. It would thus be 
important to investigate the polarization using alternative methods. 
We verified that the exponent of the power-law suppression of 
$G_{\rm total}(\delta)$ is independent of $\gamma B_x$ and thus to leading order in $U$ given by Eq.~(\ref{beta})
with $t$ replaced by $t_{\rm eff}$.

It would now be very interesting to investigate the scaling of the conductance and the dependence 
of the polarization on the length of the quantum wire for the setup with a potential step.\cite{streda}. 
Unfortunately, 
it turns out that the relevant $2 k_F$-scattering component of the potential step  
which is turned on adiabatically at the left contact but is turned off abruptly in the middle of the wire
with a step height of a few ten  percent of the band width
is fairly small. We thus 
cannot reach the scaling regime for the accessible system sizes of order $10^3$ to $10^4$. For larger step 
height the physics is dominated by ``trivial'' band effects we are not interested in. As for weak (local) 
single impurities the total conductance in the presence of the potential step decreases weakly with increasing 
$N$ (no power-law scaling at the corresponding $\delta$) while the components of the polarization oscillate with 
$N$ with an amplitude which is constant. One can enhance the effect of the potential step (the size of the 
$2 k_F$-scattering component) by adding a (sufficiently strong) single impurity located in the middle of the 
system. For this combined inhomogeneity we observe exactly the same behavior as shown in Fig.~\ref{pollarge} for 
a pure (strong) single impurity. Having this in mind we conjecture that a potential step with a sufficient 
large step size not limited by the finite band width will have the same effect on the total conductance 
(power-law suppression with length of wire) and the spin polarization (oscillation with an 
amplitude independent of the wire length) as a single localized impurity.

\section{Summary} 
\label{summary}

We have investigated the effect of SOI, a magnetic field and the Coulomb interaction on the 
transport properties of a 1D quantum wire attached to two semi-infinite noninteracting and nonmagnetic 
leads. Motivated 
by analytical calculations for a lead-less, noninteracting system described by a continuum model,\cite{streda,birkholz1} 
we have constructed a corresponding lattice model. The combined effect of SOI and a magnetic field 
led to a spin polarization  
which could be varied over a wide range in the presence of inhomogeneities (single impurity, potential step) 
by adjusting the energy of the incoming electrons (chemical potential). 
 
Using the fRG, we were able to include the Coulomb interaction in our system. We 
distinguished the cases of short quantum wires (of the order of a few tens of nanometers) and long ones 
(scaling regime; a few hundred nanometers). For short wires, we showed that the energy regime for 
which spin polarization can be found strongly depends on the Coulomb interaction and might even become 
very small depending on the other system parameters. For long wires the well-known power-law suppression of
the total conductance of an inhomogeneous Luttinger liquid as a function of the system size was 
obtained with a scaling exponent which 
depends on the two-particle interaction, the filling (both as for vanishing SOI), but also on the strength 
of the SOI via an effective nearest-neighbor hopping. However, the spin polarization as a function of system size 
exhibited oscillations with constant amplitude, not signaling any suppression on low energy-scales 
(that is for large system sizes). We found indications that the amplitude of the oscillations and 
thus the degree of spin-polarization, might even become larger than for vanishing Coulomb interaction.     

Our results indicate the importance of the two-particle Coulomb interaction in the spin filter 
suggested in Ref.~\onlinecite{streda}. In most studies on spintronic devices these correlations are 
neglected even if the suggested setups contain 1D quantum wires. Obtaining a deeper understanding of 
the performance of 1D spin filters within more realistic models of interacting electrons presents a
challenge for future theoretical studies.


\begin{thebibliography}{*}
\bibitem{remarkfootnote} The SOI parameters are given by $\alpha_i=eE_i/(4mc^2)$, $i=y,z$, 
with $e$ being the (negative) electron charge and $m$ the effective mass.\cite{birkholz1} The electric field 
$E_i$ as the gradient of the confining scalar potential is chosen to be negative here. 
Therefore, $\alpha_y,\alpha_z>0$.
\bibitem{enss1} The details of Feshbach projection for a similar problem using the same 
notation are given in Ref.~\onlinecite{enss}.
\bibitem{footfoot} A reader not interested in how we compute the interactind Green's function entering 
the generalized Landauer-B\"uttiker formula might skip this subsection.
\bibitem{footy} We note that our approximate treatment of the two-particle interaction does not capture the 
Mott insulator state. As long as we stay away from half filling for long wires this is irrelevant 
in the present context.
\end{thebibliography}
\end{document}